# A folding inhibitor of the HIV-1 Protease

R. A. Broglia[123], D. Provasi[12], F. Vasile[4],
G. Ottolina[5], R. Longhi[5], G. Tiana[12]

1) Dipartimento di Fisica, Università di Milano,
via Celoria 16, 20133 Milano, Italy
2) INFN, Sez. di Milano, Milano, Italy and
3) Niels Bohr Institute, University of Copenhagen,
Bledgamsvej 17, 2100 Copenhagen, Denmark
4) Dipartimento di Scienze Molecolari Agroalimentari,
University of Milano, via Celoria 2, Milano, Italy
5) Istituto di Chimica del Riconoscimento Molecolare,
CNR, via Bianco 21, Milano, Italy

Corresponding author: G. Tiana, Dept. of Physics, via Celoria 16, 20133 Milano, Italy; tel. +39-0250317221; fax +39-0250317487

# Abstract

Being the HIV-1 Protease (HIV-1-PR) an essential enzyme in the viral life cycle, its inhibition can control AIDS. The folding of single domain proteins, like each of the monomers forming the HIV-1-PR homodimer, is controlled by local elementary structures (LES, folding units stabilized by strongly interacting, highly conserved, as a rule hydrophobic, amino acids). These LES have evolved over myriad of generations to recognize and strongly attract each other, so as to make the protein fold fast and be stable in its native conformation. Consequently, peptides displaying a sequence identical to those segments of the monomers associated with LES are expected to act as competitive inhibitors and thus destabilize the native structure of the enzyme. These inhibitors are unlikely to lead to escape mutants as they bind to the protease monomers through highly conserved amino acids which play an essential role in the folding process. The properties of one of the most promising inhibitors of the folding of the HIV-1-PR monomers found among these peptides is demonstrated with the help of spectrophotometric assays and CD spectroscopy.

**INTRODUCTION**

HIV–1–PR is a homo-dimer, that is a protein whose native conformation is built of two (identical) disjoint chains (see Fig. 1) each of them made of 99 amino acids. Sedimentation equilibrium experiments have shown that at neutral pH the protease folds according to a three–state mechanism ($2U \rightarrow 2N \rightarrow N_2$), populating consistently the monomeric native conformation N. The dimer dissociation constant is $K_D = 5.8$ μM at 4°C, while the folding temperature of the monomer, i.e. the temperature at which the free energy of the native monomeric state N is equal to that of the unfolded state U is $T_f = 52.5$°C [1]. Some recent NMR studies have also found folded monomers of several mutants[2-4]. At low pH, on the other hand, calorimetric experiments have shown[5] that there is a single transition at T=59°C (pH 3.4, 25 μM protein, 100mM NaCl) between the dimeric native state and a monomeric unfolded state.

The pH of the solution which surrounds the HIV-1-PR affects its catalytic capabilities. It has been shown[6] that the activity of the protease increases if the pH of the solution is lowered. Considering that the active site of the protease is at the interface between the two monomeric units, it is most likely that the affinity of the protease to the substrate is correlated with the structure of the dimer. The value of the dimerization constant $K_D$ is much controversial[1]. Anyway it seems it ranges from the order of nM to that of μM in going from acidic to neutral conditions. One can thus guess that the activity of the protease is higher at low pH because the dimeric state is more populated. Increasing the value of pH, acid residues acquire a negative charge. In particular, the pair of ASP25 which lie close on the interface repel each other through the Coulomb force. The overall effect is to increase the dissociation constant (measured by sedimentation equilibrium experiments) which assumes the value $K_D$=5.8 µM at pH 7 (and T=4°C [1]), further increasing at higher temperatures. Consequently, one expects a detectable ratio of folded monomers in solution. Thus, the destabilization of the monomer will lead to enzyme inhibition[7].

Drug resistance has severely limited the effectiveness of conventional (active-site centered) HIV-1 protease inhibitors in AIDS therapy[8]. Experimental evidence has shown that drug-resistant mutations can occur only at specific positions that are critical for drug binding but are tolerated as far as folding and thus viral activity is concerned. Likely resistance-evading drugs can thus be searched among molecules interacting strongly with those conserved residues which play an important role in the folding of the protease. Following this viewpoint, we have recently proposed a general strategy based on the inhibition of folding [7].

Model studies of single domain globular proteins[9-11] indicate that folding proceeds following a hierarchical succession of events starting from the formation of local elementary structures (LES, stabilized by strongly interacting, highly conserved, as a rule hydrophobic, so called "hot" amino acids). The docking of these LES, which is again controlled by these "hot" amino acids, leads to the formation of the (post-critical) folding nucleus[12]. Mutations of the "hot" amino acids give rise, in general, to protein denaturation [13]. Strong support for the soundness of this hierarchical scenario is found in a number of circumstantial evidences[14-21].

The same scenario applies to each of the monomers forming three-state folding homo-dimers, like the HIV-1-PR[22], as has been shown in detail with the help of extensive Go-model simulations[7] (see also the detailed all-atom Go model simulations of ref. 23).

The hierarchical model also suggests that it is possible to destabilize the native conformation of a protein with the help of peptides whose sequences are identical to those of the LES of the protein[24]. Such peptides (p–LES) interact with the protein (in particular with their complementary LES) with the same energy which stabilizes its folding nucleus, thus competing with its formation. Given this fact, it is unlikely that the virus can develop drug–resistance through mutations. This is because, to prevent interaction between the LES or between the p-LES and the LES, one has to mutate "hot" amino acids.

Based on these general criteria, molecular dynamics and Monte Carlo simulations, along with evidence taken from site-directed mutagenesis and sequence analysis, lead to the identification of three segments of the HIV-1-PR monomers which are likely connected with the LES of the protease[7]. The segment associated with the stretch 83-93 of the protease is expected to be the most promising candidate as monomer inhibitor. In the following we will show its efficiency as inhibitor of the enzyme.

MATERIALS AND METHODS

Recombinant **HIV-1-PR**, expressed in E. Coli (Bachem UK, Ltd., catalog H-9040)(25,26) contained five mutations to restrict autoproteolysis (Q7K, L33I, L36I) and to restrict cysteine thiol oxidation (C67A and C95A). The enzyme was stored at (-70ºC) as solution with concentration 0.1 µg/µL in dilute HCl, (pH=1.6).

A **chromogenic substrate** for HIV-1-PR (HIV Protease Substrate III, Bachem UK Ltd., catalog H-9035) with sequence H-His-Lys-Ala-Arg-Val-Leu-Phe(NO2)-Phe-Glu-Ala-Nle-Ser-NH$_2$ was obtained as a 1 mg desiccate, diluted with 0.1 ml of DMSO, and stored at -20ºC. Protease-assisted cleavage between the Leu and the Phe(NO$_2$) residues of substrate entails a blue-shift of the absorption maximum (277 nm to 272 nm) that can be adequately monitored observing the continuous decrease of absorbance at 300 nm [27-29]. A regression of the absorbance at 300 nm against substrate concentration allows to check that the absorbance scales linearly up to a concentration of 800 µM, and to estimate the molar absorption coefficient of the whole substrate ($\varepsilon_S = 3000 \pm 600$ (M cm)$^{-1}$). To determine the molar absorption coefficient of the cleaved products, reactions with different initial substrate concentration were followed for at least 2 hours. The absorbance at 300 nm after complete peptidolysis allows to determine differential extinction coefficient $\Delta\varepsilon = 500 \pm 90$ (M cm)$^{-1}$ between the whole substrate and the cleaved products. This compares well with a difference of extinction coefficient at 310 nm between the cleaved and the complete substrate of $1200 \pm 100$ $(M\ cm)^{-1}$, reported in ref. 27 for a similar substrate.

**Inhibitor peptide** (peptide **I**, cf. Table 1) from the sequence of the HIV–1–PR wild type (PDB code 1BVG) were synthesized by Fmoc solid-phase peptide synthesis with acetyl and amide as terminal protection group and was estimated to be > 95% pure by analytical HPLC after purification. After that 1 mg of inhibitor peptide was dissolved in 100 µl of DMSO, 4 µl of this solution were then diluted with 16 µl of DMSO and 180 µl of the buffer used for assay. The obtained solution (150 µM of peptide **I**) was used for the experiments.

**Control peptides** were also synthesized by Fmoc solid-phase peptide synthesis. Two of them, called $K_1$ and $K_2$ (cf. Table 1) are also form the primary sequence of the HIV–1–PR, but from regions well outside the local elementary structures identified in ref. 7. A third peptide $K_3$ is not related in any way to the protease. It is to be noted that peptide $K_2$ is rather hydrophobic and only > 70% purity could be achieved.

The **assay buffer** was prepared, following ref. 27 by adding 0.8 mM NaCl, 1 mM EDTA and 1 mM dithiothreitol to a 20 mM phosphate buffer (pH 6).

**Experimental methods**

Each measure was performed recording the absorbance at 300 nm in a standard UV-vis spectrophotometer (Jasco V-560). The sample had a total volume of 70 µL in Spectrosil Far UV Quartz (170-2700 nm) cuvettes (3.3 mm optical path). The sample in the cuvette was exposed to a constant temperature (25 ± 0.05°C) provided by continuous circulation of water from a water bath to the cell holder via a circulation pump.

For the determination of the kinetic parameters, we measured at least 6 different concentration of substrate, spanning the range from 100 µM to 600 µM. After proper thermal stabilization of the substrate dissolved in the buffer, the absorbance at 300 nm was checked to be stable, then the reaction initiated by adding 4 µl of the enzyme solution.

For each sample we followed at least 1200 seconds and determined the initial rate $v_i$ by a linear fit of the first 200 sec. We repeated twice the determination of the kinetic parameters of the enzyme.

The assay was then repeated in presence of the inhibitor peptide. We followed the same procedure, incubating for 60 sec. the inhibitor peptide (3, 10 and 20 µM) with the protein before adding it to the substrate.

**Circular dichroism spectrum**

Ultraviolet CD spectra were recorded on a Jasco J-810 spectropolarimeter in nitrogen atmosphere at room temperature using 0.1 cm path-length quartz cell. Each spectrum was recorded between 260-200 nm. The data were collected at a rate of 10 nm/min with a wavelength step of 0.2 nm and a time constant of 2 sec. The spectra were corrected with respect to the baseline and normalized to the aminoacidic concentration. The protein and the peptide were dissolved in a 20 mM phosphate buffer with 0.8 M NaCl at the same concentration used for the activity assays. The CD spectra were analyzed in terms of contribution of secondary structure elements[30] using the K2D method based on comparison with CD spectra of proteins and peptides with known secondary structure.

**RESULTS AND DISCUSSION**

**Enzyme kinetics and inhibition constants** are analyzed in the framework of the Michaelis-Menten equation. That is, one assumes that the reaction can be described by the relation

$$E + S \leftrightarrow ES \rightarrow E + P, \qquad (1)$$

where E, S, ES and P stand for enzyme, substrate, enzyme-substrate complex and product, respectively. The rate in the production of the product [P] for short times, is then described by

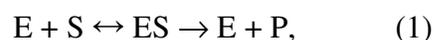

$$v_i = v_{max} [S] (K_m + [S])^{-1}, \qquad (2)$$

where $K_m$ represents the dissociation constant of the enzyme substrate complex, and $v_{max} = k_{cat} [E]_0$ is the maximum catalytic rate, attained for saturating substrate concentrations [S]. The quantity $k_{cat}$ is the catalytic constant, i.e. the first order rate constant for the chemical conversion of the ES complex into the EP complex. The value of $K_m$ of the enzyme-substrate can be obtained by transforming the reaction rate and concentration data to a double-reciprocal plot (see Fig. 2).

A fit of the data provides estimates for $K_m$ and $v_{max}$ (380±80 µM and 8.57±0.88 mAbs/min, respectively). We note that this value of $K_m$ is considerably larger than the one reported in the literature for the same enzyme-substrate system. It should be stressed, however, that the experimental conditions in ref. 29 are different, as far as pH and ionic strength are concerned.

In presence of an inhibitor, Eq. 2 still holds, where now the parameters $v_{max}$ represent the apparent maximum catalytic rate for the inhibited reaction, and $K_m$ should be interpreted as an apparent dissociation constant. To test the inhibitory properties of Peptide **I**, we have measured these kinetic parameters $K_m^{app}$ and $v_{max}$ for three different concentrations of the peptide. The data recorded are plotted in Figure 2, along with the data obtained without inhibitors. The results of the fits are also reported in Table 2.

It is observed that the values of $v_{max}$ are almost constant and only the dissociation constants $K_m$ increase with increasing concentration of peptide **I**. The observed kintetics is thus compatible with a competitive inhibition mechanism, where

$$K_m^{app} v_{max}^{-1} = K_m v_{max}^{-1} ( 1 + [I] K_i^{-1}). \quad (3)$$

Using this equation, we can estimate from a linear regression of the slopes of the Lineweaver-Burk plot, the value $K_i$=2.58±0.78 µM, (cf. inset to Fig. 2) that gives the dissociation constant for the enzyme-inhibitor complex. The results are reported in Table 2. It should be stressed that, due to the saturation of the linear relationship between absorbance and concentration at around 800 µM, the most concentrated sample is only twice $K_m$, limiting the numerical accuracy of the estimated parameters. On the other hand the data clearly points to a competitive inhibition mechanism, where the binding of peptide **I** to the protein, causing its unfolding, competes with its folded, active conformation.

We have made similar measurements using the **control peptides** $K_1$, $K_2$ and $K_3$ instead of peptide **I**, and found no appreciable variation in the kinetic parameters with respect to the uninhibited case (see Fig. 3). In presence of peptide $K_1$, $K_2$ or $K_3$ the reaction displayed initial rates $v_i$ = 1.98 mAbs/min, $v_i$ = 2.06 mAbs/min, and $v_i$ = 2.01 mAbs/min, respectively, essentially identical to the value of $v_i$ obtained for the uninhibited enzyme in identical conditions. Due to the limited amount of enzyme available, we have not fully characterized the kinetics of the reaction in presence of the control peptides, but just checked that for a given value of [S], the reactions were not affected.

To provide evidence demonstrating that the inhibition mechanism of the peptide indeed prevents the proper folding of the enzyme, we measured a **circular dichroism** spectrum of the protein alone and after incubation with the peptide. The CD spectrum of the protease (cf. Fig. 4) under the same conditions used for the activity assay indicate a beta-sheet content of 30%, consistent with the beta character of the native conformation[31]. Figure 4 also displays the CD spectrum of the solution of the protease plus the **I** inhibitor (to which the spectrum of the Peptide **I** alone has been subtracted) at the same concentrations and under the same conditions as those of the activity assay. It shows a loss of beta–structure (to a beta–sheet content value of 14%), indicating that the protein is, to a large extent, in a non-folded conformation. These numbers compare well with those predicted by model calculations[7].

**CONCLUSIONS**

The peptide **I** (≡ NIIGRNLLTQI) displaying a sequence identical to that of one of the LES (83-93) of each of the two identical chains forming the HIV–1–PR homodimer is found to be a highly specific and efficient inhibitor ($K_i$ = 2.58 ± 0.78 µM) of the folding of the 99mers, and thus of the whole enzyme. A remarkable property of this inhibitor is that it is unlikely that it would allow for escape mutants. In fact, the only mutations which will prevent **I** from acting are likely to involve protein denaturation.

Obvious disadvantages of the inhibitor are its length, its hydrophobicity and its peptidic character, as it is not clear how to prevent the degradation by enzymes. Consequently, there are a number of clear tasks lying ahead in the quest to develop the lead into a potential drug. One is to investigate whether the shortening of Peptide **I**, by leaving out some residues either at the beginning or at the end (or both), lead to peptides which still inhibit folding with similar specificity and effectiveness as Peptide **I** does and, at the same time, are more soluble. Another is to develop molecules mimetic to the present inhibitor or eventually to shorter peptides derived from it.

**A**CKNOWLEDGEMENTS

The support and advice of G. Carrea and G. Colombo (CNR, Institute of Chemistry of Molecular Recognition, Milano) is gratefully acknowledged. We wish also to thank R. Jennings, F. Garlaschi, G. Zucchelli (University of Milano, Biology depertment) and E. Ragg (University of Milano, Dipartimento di Scienze Molecolari Agroalimentari) for help and discussions.

| Peptide | HIV1- Pr sites | Sequence | Molecular weight |
|---|---|---|---|
| **I** | 83–93 | NIIGRNLLTQI | 1295.54 |
| $K_1$ | 61-70 | QILIEICGHK | 1194.46 |
| $K_2$ | 9-19 | PLVTIKIGGQL | 1179.46 |
| $K_3$ | not from HIV-1-PR | LSQETFDLWKLLPEN | 1874.12 |

**Table 1**. **Peptides used in this work.**
Peptide **I** is the proposed inhibitor; control peptides, either from the HIV-1-PR sequence ($K_1$ and $K_2$) and not ($K_3$) were also tested.

|  | No inhibitor | [**I**]=3 µM | [**I**]=10 µM | [**I**]=20 µM |
|---|---|---|---|---|
| $K_m$ (µM) | 380±80 | 680±92 | 980±290 | 2600±2000 |
| $v_{max}$ (mAbs/min) | 8.57±0.88 | 10.37±0.8 | 9.29±1.9 | 9.71±11 |
| $v_{max}$ (µmol/s) | 0.94±0.19 | 1.14±0.09 | 1.02±0.27 | 1.07±1.2 |

**Table 2. Kinetic parameters for the reactions assessed**. The table contains the kinetic data for the HIV-1-PR hydrolization of the HIV-1-PR substrate III without and with different concentrations [**I**] of inhibitor. The values of $K_m$ for the inhibited reactions have to be regarded as apparent dissociation constants. The values for $v_{max}$ are reported as measured (mAbs/min) and converted using the differential extinction coefficient $\Delta\varepsilon=500 \pm 90$ (M cm)$^{-1}$. Values and errors are obtained using non-linear fits.

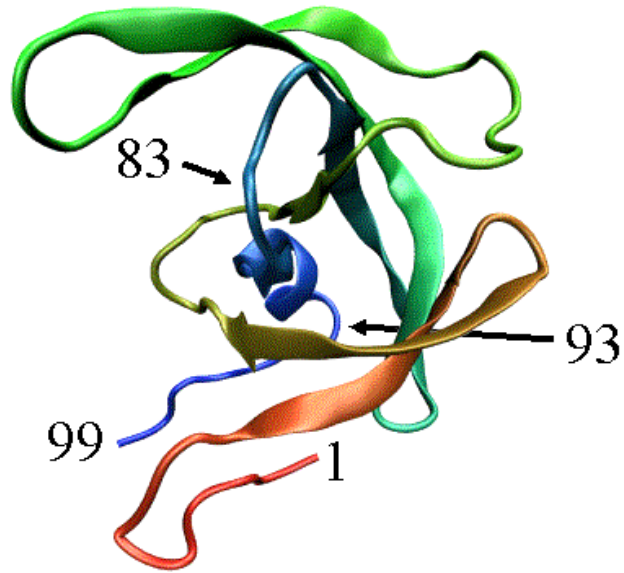

**Figure 1: Monomer of the HIV1-Protease.** The inhibitor peptide **I** has a sequence identical to that of the segment whose ends (residue 83 and 93) are indicated in the figure.

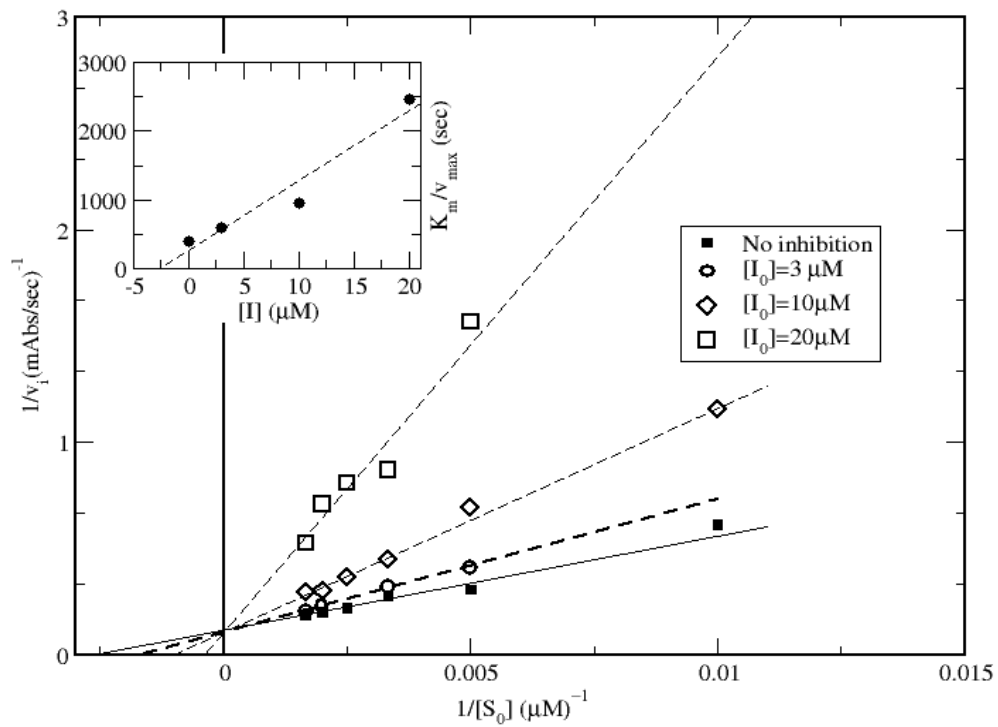

**Figure 2: Inhibitory activity of Peptide.** The Lineweaver-Burk plot associated with the protease (filled squares) and the protease complexed with the inhibitor **I** at 3 μM (open circles), 10 μM (open diamonds) and 20 μM (open squares). The lines are the fits to the experimental points. The initial velocities $v_i$ are expressed in terms of μM/s, while the substrate concentration $[S_0]$ is in μM. In the inset we report the values of $K_m/v_{max}$ as a function of the inhibitor concentration [I]. The linear fit to the data gives a $K_i=2.58\pm0.78$ μM, with a correlation coefficient r=0.94 and a p-value for the F-test of 0.029.

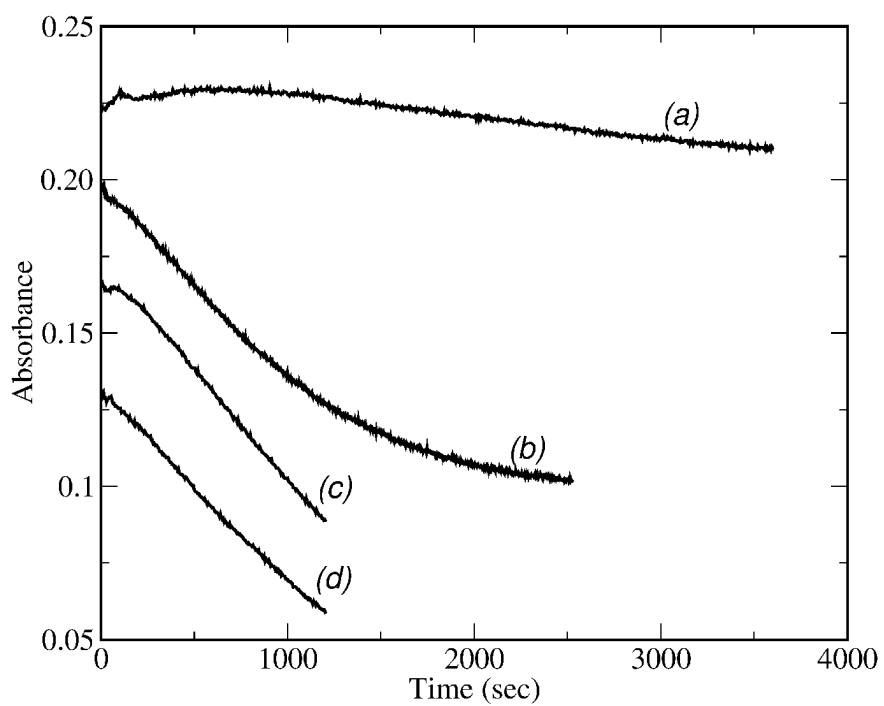

**Figure 3: Control peptides.** The enzymatic kinetics of the inhibited protease alone (curve *b*), of the protease inhibited with peptide **I** (*a*), and of the protease together with control peptides $K_1$ (*c*) and $K_2$ (*d*), measured as change in absorbance of the chromogenic substrate as a function of time. All the curves have been measured at $[S_0] = 125\mu M$, and have been shifted along the y-axis in order to be easily inspected. In all samples the concentration of peptide was 3 µM.

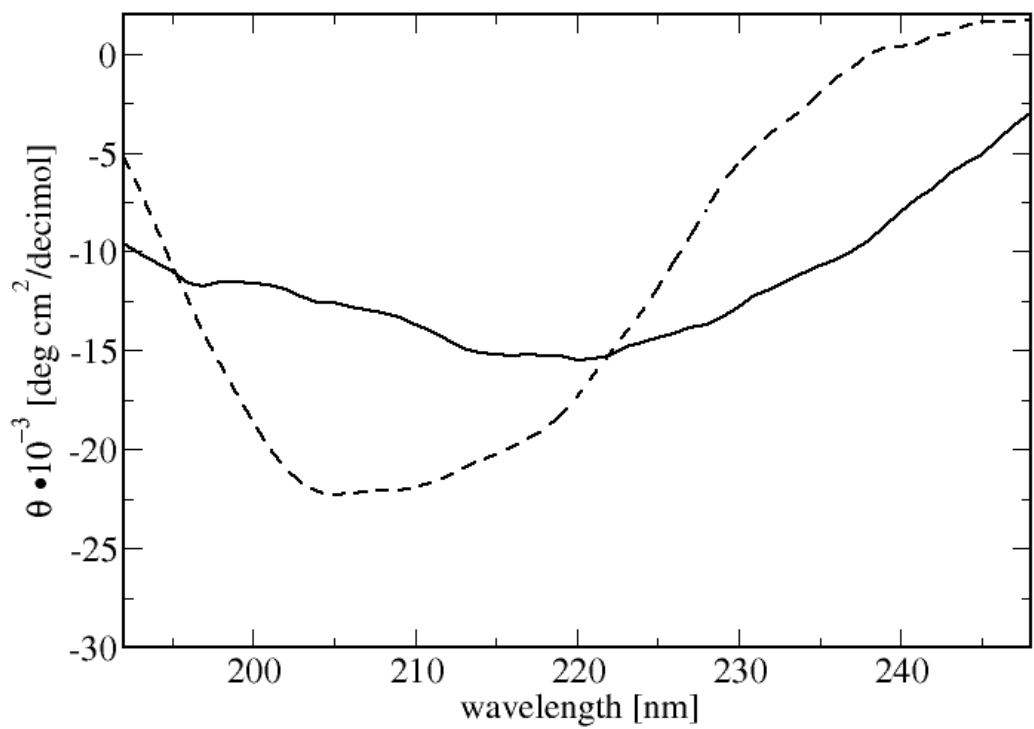

**Figure 4: Evidence of unfolding.** The circular dichroism spectrum of the protease (dashed curve)
and of the solution composed of the protease and peptide **I** (continuous curve) in the ratio 1:3, from which the spectrum of the peptide has been subtracted.